\documentclass[conference,a4paper]{IEEEtran}
\IEEEoverridecommandlockouts

\usepackage[hidelinks]{hyperref}
\usepackage{nameref}
\usepackage[cmex10]{amsmath}
\usepackage{amssymb,amsfonts}
\usepackage{dblfloatfix}

\usepackage[ruled,vlined]{algorithm2e}
\usepackage{graphicx}
\usepackage{subcaption}

\usepackage{booktabs}
\usepackage{siunitx}
\usepackage[numbers,compress]{natbib}
\usepackage{texnames}
\usepackage{bm,bbm}
\usepackage{orcidlink}

\usepackage{float}

\usepackage{tikz}
\usetikzlibrary{quantikz2}
\usetikzlibrary{shapes.geometric, arrows}

\tikzstyle{block} = [rectangle, draw, fill={rgb,255:red,23;green,190;blue,207}, fill opacity=0.3, text opacity=1, text width=12em, text centered, rounded corners, minimum height=2em]
\tikzstyle{doubleblock} = [rectangle, draw, fill={rgb,255:red,23;green,190;blue,207}, fill opacity=0.3, text opacity=1, text width=12em, text centered, rounded corners, minimum height=3em]
\tikzstyle{arrow} = [thick, ->, >=stealth]

\usepackage{makecell}
\usepackage{multirow}
\usepackage{colortbl}

\usepackage{mathtools}

\usepackage[most]{tcolorbox}

\usepackage{url}
\usepackage[font=small]{caption}

\newcommand{\TrainingSet}{\bm{T}} 
\newcommand{\TestSet}{\Psi}

\newcommand{\GT}{\rm GT}

\newcommand{\TP}{\rm TP}
\newcommand{\FP}{\rm FP}
\newcommand{\FN}{\rm FN}
\newcommand{\TN}{\rm TN}

\newcommand{\ACC}{\rm ACC}

\newcommand{\SEN}{\rm SEN}
\newcommand{\SPE}{\rm SPE}
\newcommand{\Fscore}{\rm F1}
\newcommand{\MCC}{\rm MCC}

\definecolor{textred}{rgb}{0.8431, 0.1608, 0.0588}
\definecolor{textgreen}{rgb}{0.4588, 0.7451, 0.3059}
\definecolor{textyellow}{rgb}{0.9961, 0.7608, 0.0512}

\definecolor{celllightred}{rgb}{0.988, 0.761, 0.761}
\definecolor{celllightgreen}{rgb}{0.79216, 0.89804, 0.72941}
\definecolor{celllightyellow}{rgb}{1, 0.90588, 0.68235}
\definecolor{celllightblue}{rgb}{0.608, 0.741, 0.996}

\usepackage{wasysym}
\newcommand{\cmark}{\CIRCLE}%
\newcommand{\xmark}{\Circle}

\begin{document}

\title{\uppercase{Light-cone feature selection in methane hyperspectral images}
\thanks{AM was supported by the Priority Research Areas Anthropocene and Digiworld under the program Excellence Initiative – Research University at the Jagiellonian University in Kraków. JN and AMW were supported by the Silesian University of Technology funds through the grant for maintaining and developing research potential. AMW also received funding for research from the National Science Centre (project: DEC-2024/08/X/ST6/01777). }
}

\author{	\IEEEauthorblockN{Artur\ Miroszewski\orcidlink{0000-0002-0391-8445}}
	\IEEEauthorblockA{\textit{Institute of Theoretical Physics},\\
        \textit{Mark Kac Center for Complex}\\ \textit{Systems Research}\\
		Jagiellonian University\\
        Łojasiewicza 11, 30-348 Cracow, Poland\\
		artur.miroszewski@uj.edu.pl}
	\and
        \IEEEauthorblockN{Jakub\ Nalepa\orcidlink{0000-0002-4026-1569}}
	\IEEEauthorblockA{\textit{Silesian University of Technology} \\ Akademicka 2A, 44-100 Gliwice\\\textit{KP Labs}\\
        Bojkowska 37J, 44-100 Gliwice,\\Poland\\
		jnalepa@ieee.org}
        \and
	\IEEEauthorblockN{Agata\ M. Wijata\orcidlink{0000-0001-6180-9979}}
	\IEEEauthorblockA{\textit{Silesian University of Technology} \\ Akademicka 2A, 44-100 Gliwice\\\textit{KP Labs}\\
        Bojkowska 37J, 44-100 Gliwice,\\Poland\\
		awijata@ieee.org}
}

\maketitle
\begin{abstract}
    Hyperspectral images (HSIs) capture detailed spectral information across numerous contiguous bands, enabling the extraction of intrinsic characteristics of scanned objects and areas. This study focuses on the application of light-cone feature selection in quantum machine learning for methane detection and localization using HSIs. The proposed method leverages quantum methods to enhance feature selection and classification accuracy. The dataset used includes HSIs collected by the AVIRIS-NG instrument captured in geographically diverse locations. In this study, we investigate the performance of support vector machine classifiers with different classic and quantum kernels. The results indicate that the quantum kernel classifier, combined with light-cone feature selection, provides in one metric, superior performance when compared to the classic techniques. It demonstrates the potential of quantum machine learning in improving the remote sensing data analysis for environmental monitoring.
\end{abstract}

\begin{IEEEkeywords}
	Quantum Machine Learning, feature selection, methane detection, classification, Earth observation.
\end{IEEEkeywords}

\section{Introduction}

Hyperspectral images (HSIs) capture up to hundreds of contiguous and narrow spectral bands within the electromagnetic spectrum. Such detailed imagery allows for extracting intrinsic characteristics of the scanned objects and areas, if HSIs are remotely sensed from a drone, aircraft or an imaging satellite~\cite{10180082}. There are a multitude of Earth observation use cases which may directly benefit from such images, and they span---but are not limited to---precision agriculture, environmental monitoring, event and disaster detection, change detection and tracking, surveillance, and many more~\cite{7882742}. One of the recently investigated applications involving the analysis of remotely-sensed HSIs is methane detection, as detecting and monitoring this greenhouse gas play a pivotal role in environmental monitoring~\cite{10281786,PEI2023113652,Ruzicka_starcop_2023}. If (super)emitters are effectively detected, appropriate actions may be tackled in a timely fashion~\cite{2022_Abbass}.

However, although HSIs offer very detailed information about the scanned areas, their transfer, storage and analysis pose lots of practical challenges, as such images may become extremely large, especially when they are captured on board a satellite. Here, in-orbit data acquisition delivers benefits in terms of spatial scalability, but the transfer of HSIs to the ground may easily become infeasible due to the downlink constraints. Therefore, the current trend in the field is to bring the analysis chain on board a satellite to extract actionable items and insights before sending them back to Earth~\cite{10180082}. Additionally, it is worth emphasizing that---for specific applications---only a subset of all spectral bands (and possibly other features extracted from HSIs) play a key role. Therefore, dimensionality reduction is of paramount importance in both in-orbit and on-the-ground HSI analysis, as it can reduce the computational, storage, and transfer requirements. We follow the path of determining the most important subsets of spectral bands for an important downstream task of methane detection and localization in hyperspectral images using a quantum machine learning-powered approach---quantum algorithms have been recently investigated in Earth observation due to their possible quantum advantage~\cite{10214376, delilbasic2021quantum}.

\subsection{Contribution}
\label{sec:contribution}
We introduce a novel approach to feature selection for HSIs using light-cone feature selection \cite{suzuki2024light} within a quantum machine learning framework. Our contributions are as follows:
\begin{itemize}
\item \textbf{Quantum Kernel Classifier}---we propose a quantum kernel classifier based on local projected kernels, which allows for the investigation of the light-cone feature selection method.

\item \textbf{Methane Detection and Localization}---we apply our quantum kernel classifier to the task of methane detection and localization, a critical application in environmental monitoring. By utilizing the STARCOP dataset, we demonstrate the effectiveness of our approach in identifying methane emissions from hyperspectral data.

\item \textbf{Comparative Analysis}---we conduct a comprehensive comparative analysis of different support vector machine (SVM) classifiers, including linear, radial basis function (RBF) and quantum kernels. We validate our methods using standard metrics and statistical tests, ensuring the robustness and reliability of our results. 

\item \textbf{Feature Importance}---we analyze the importance of different spectral bands and the mag1c enhancement map in the context of methane detection. 

\end{itemize}

This study demonstrates the potential of quantum machine learning in advancing the field of remote sensing and environmental monitoring, providing a foundation for future research and applications.

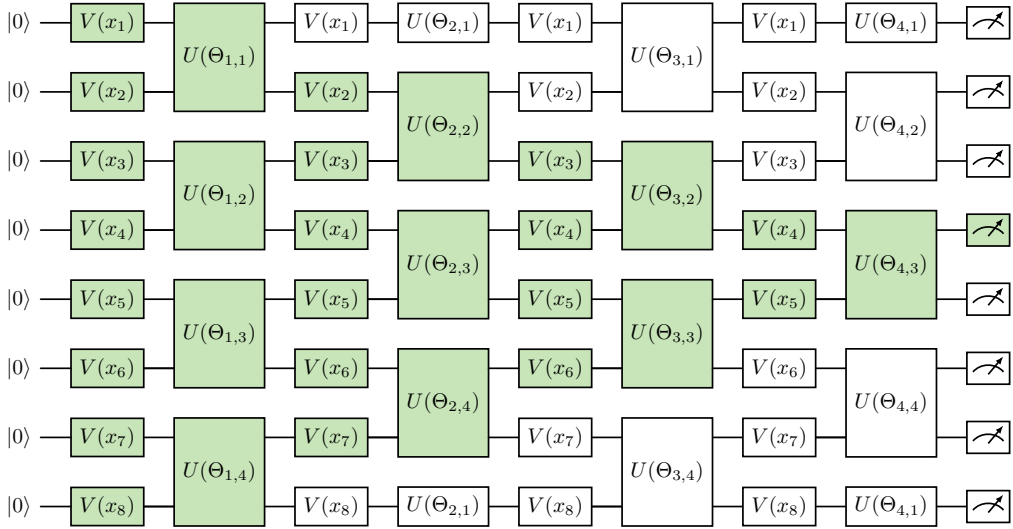
\begin{figure*}
    \centering

\scalebox{0.8}{
\begin{quantikz}
\lstick{$\ket{0}$} & \gate[style={fill={rgb,1:red,0.4588;green,0.7451;blue,0.3059},fill opacity=0.4}]{V(x_1)} & \gate[2, style={fill={rgb,1:red,0.4588;green,0.7451;blue,0.3059},fill opacity=0.4}]{U(\Theta_{1,1})} &  \gate{V(x_1)} & \gate{U(\Theta_{2,1})} & \gate{V(x_1)} & \gate[2]{U(\Theta_{3,1})} & \gate{V(x_1)} & \gate{U(\Theta_{4,1})} & \meter{} \\
\lstick{$\ket{0}$} & \gate[style={fill={rgb,1:red,0.4588;green,0.7451;blue,0.3059},fill opacity=0.4}]{V(x_2)} & \qw &  \gate[style={fill={rgb,1:red,0.4588;green,0.7451;blue,0.3059},fill opacity=0.4}]{V(x_2)} & \gate[2, style={fill={rgb,1:red,0.4588;green,0.7451;blue,0.3059},fill opacity=0.4}]{U(\Theta_{2,2})} & \gate{V(x_2)} & \qw & \gate{V(x_2)} & \gate[2]{U(\Theta_{4,2})} & \meter{} \\
\lstick{$\ket{0}$} & \gate[style={fill={rgb,1:red,0.4588;green,0.7451;blue,0.3059},fill opacity=0.4}]{V(x_3)} & \gate[2, style={fill={rgb,1:red,0.4588;green,0.7451;blue,0.3059},fill opacity=0.4}]{U(\Theta_{1,2})}  & \gate[style={fill={rgb,1:red,0.4588;green,0.7451;blue,0.3059},fill opacity=0.4}]{V(x_3)} & \qw & \gate[style={fill={rgb,1:red,0.4588;green,0.7451;blue,0.3059},fill opacity=0.4}]{V(x_3)} & \gate[2, style={fill={rgb,1:red,0.4588;green,0.7451;blue,0.3059},fill opacity=0.4}]{U(\Theta_{3,2})} & \gate[]{V(x_3)} & \qw & \meter{} \\
\lstick{$\ket{0}$} & \gate[style={fill={rgb,1:red,0.4588;green,0.7451;blue,0.3059},fill opacity=0.4}]{V(x_4)} &  \qw      & \gate[style={fill={rgb,1:red,0.4588;green,0.7451;blue,0.3059},fill opacity=0.4}]{V(x_4)} & \gate[2, style={fill={rgb,1:red,0.4588;green,0.7451;blue,0.3059},fill opacity=0.4}]{U(\Theta_{2,3})}  &  \gate[style={fill={rgb,1:red,0.4588;green,0.7451;blue,0.3059},fill opacity=0.4}]{V(x_4)} & \qw & \gate[style={fill={rgb,1:red,0.4588;green,0.7451;blue,0.3059},fill opacity=0.4}]{V(x_4)} & \gate[2, style={fill={rgb,1:red,0.4588;green,0.7451;blue,0.3059},fill opacity=0.4}]{U(\Theta_{4,3})} & \meter[style={fill={rgb,1:red,0.4588;green,0.7451;blue,0.3059},fill opacity=0.4}]{} \\
\lstick{$\ket{0}$} & \gate[style={fill={rgb,1:red,0.4588;green,0.7451;blue,0.3059},fill opacity=0.4}]{V(x_5)} & \gate[2, style={fill={rgb,1:red,0.4588;green,0.7451;blue,0.3059},fill opacity=0.4}]{U(\Theta_{1,3})}  &  \gate[style={fill={rgb,1:red,0.4588;green,0.7451;blue,0.3059},fill opacity=0.4}]{V(x_5)} & \qw & \gate[style={fill={rgb,1:red,0.4588;green,0.7451;blue,0.3059},fill opacity=0.4}]{V(x_5)} & \gate[2, style={fill={rgb,1:red,0.4588;green,0.7451;blue,0.3059},fill opacity=0.4}]{U(\Theta_{3,3})} & \gate[style={fill={rgb,1:red,0.4588;green,0.7451;blue,0.3059},fill opacity=0.4}]{V(x_5)} & \qw & \meter{} \\
\lstick{$\ket{0}$} & \gate[style={fill={rgb,1:red,0.4588;green,0.7451;blue,0.3059},fill opacity=0.4}]{V(x_6)} & \qw      & \gate[style={fill={rgb,1:red,0.4588;green,0.7451;blue,0.3059},fill opacity=0.4}]{V(x_6)} & \gate[2, style={fill={rgb,1:red,0.4588;green,0.7451;blue,0.3059},fill opacity=0.4}]{U(\Theta_{2,4})} & \gate[style={fill={rgb,1:red,0.4588;green,0.7451;blue,0.3059},fill opacity=0.4}]{V(x_6)} & \qw      & \gate{V(x_6)} & \gate[2]{U(\Theta_{4,4})} & \meter{} \\
\lstick{$\ket{0}$} & \gate[style={fill={rgb,1:red,0.4588;green,0.7451;blue,0.3059},fill opacity=0.4}]{V(x_7)} & \gate[2, style={fill={rgb,1:red,0.4588;green,0.7451;blue,0.3059},fill opacity=0.4}]{U(\Theta_{1,4})} & \gate[style={fill={rgb,1:red,0.4588;green,0.7451;blue,0.3059},fill opacity=0.4}]{V(x_7)} & \qw & \gate{V(x_7)} & \gate[2]{U(\Theta_{3,4})} & \gate{V(x_7)} & \qw & \meter{} \\
\lstick{$\ket{0}$} & \gate[style={fill={rgb,1:red,0.4588;green,0.7451;blue,0.3059},fill opacity=0.4}]{V(x_8)} & \qw & \gate{V(x_8)} & \gate[]{U(\Theta_{2,1})} & \gate[]{V(x_8)} & \qw & \gate{V(x_8)} & \gate{U(\Theta_{4,1})} & \meter{} \\
\end{quantikz}
}

    \caption{Schematic representation of the light-cone feature selection circuit used in the simulations, $n=8$, $L=4$. It consists of the alternating layers of one-qubit and two-qubit gates. One qubit gates, $V(x_i)$ are responsible for the re-uploading of the $i^{th}$ feature of the data point $x$. Two-qubit blocks $U(\Theta_{k,l})$ are responsible for the information propagation of the features throughout the qubits. Each qubit is measured at the end and gives rise to the local, projected kernel. With green we highlight the measurement of the fourth qubit and its light-cone.
    The weights $w_i(x_j)$ for the importance score for the highlighted local kernel are $w_{4} (x_1) = w_{4} (x_8) = 1,\ w_{4}(x_2) = w_{4} (x_7) = 2,\ w_{4} (x_3) = w_{4} (x_6) = 3,\ w_{4} (x_4) = w_{4} (x_5) = 4$.}
    \label{fig:circ}
\end{figure*}


\section{Materials and methods}

\subsection{Dataset}

In this study, we utilize the STARCOP dataset, designed for the detection and segmentation of methane emissions using hyperspectral images (HSI) collected by the AVIRIS-NG instrument~\cite{Ruzicka_starcop_2023}. This instrument records data in the spectral range from 400 to 2500 nm, with a spatial resolution ranging from 0.3 to 4 m. The data were collected in 2023 and cover areas related to the fossil fuel industry, where methane emissions are particularly significant. The following bands were extracted for analysis from the STARCOP data set: 1---460, 2---550, 3---640, 4---2004, 5---2109, 6---2310, 7---2350, 8---2360\,nm. The dataset consists of 3208 HSIs with dimensions $512\times 512\times 8$.

For each image, a methane enhancement map was obtained using the mag1c algorithm~\cite{2020_Foote} and a manually generated binary ground truth (\GT) image indicating areas of methane emissions was generated. The data are divided into training ($\TrainingSet$) and test ($\TestSet$) sets. Using an extended version of the Simple Linear Iterative Clustering (SLIC) algorithm (operating over all available spectral bands), we determined the superpixels for each image. This is done in order to perform sample number reduction for the subsequent analysis. The superpixel value represents the average value for each band within the analyzed area. Utilizing the knowledge of the superpixel area, we generated a label for each superpixel based on \GT: 1---\textit{methane} and 0---\textit{background}. Similarly, we determined the average methane enhancement value for each superpixel based on the available mag1c map.

Next, from the training set ($\TrainingSet$), we randomly selected 15 9-element feature vectors (the mag1c enhancement map and 8 bands) coupled with the ground-truth label corresponding to \textit{methane}, and 15 for \textit{background}. In this study, we create a new balanced test set $\TestSet'$, comprising all methane superpixels along with random background superpixels. All metrics are reported for $\TestSet'$ (with a size of $\left|\TestSet'\right|=1476$ superpixels).

\subsection{Light-cone feature selection}
In order to construct a quantum kernel classifier with the possibility of feature selection, we follow the approach of light-cone feature selection introduced in \cite{suzuki2024light}. The quantum kernel is obtained by combining local, one-qubit kernels, $\kappa_i(x,y)$'s with the corresponding weights, $\lambda_i$'s,
\begin{equation}\label{eq:kernel}
    \kappa(x,y) = \sum_{i=1}^n \lambda_i \kappa_i(x,y),
\end{equation}
where $n$ is the number of features considered, and is equal to the number of qubits used in the circuit.
Now, the value of $\lambda_i$ depends on the value of centered alignment \cite{cortes2012algorithms} of the corresponding local kernel $\kappa_i(x,y)$ and the total sum of $\lambda$'s is normalized to one. In general, the greater the value of the centered alignment of the given local kernel $\kappa_i(x,y)$ is, the greater its share in the resultant kernel value becomes. The proposed quantum embedding consists of alternating layers of one-qubit and two-qubit blocks, represented schematically in Fig. \ref{fig:circ}. The feature selection of such circuit architecture depends on the information propagation of data re-uploaded features. When we consider a local kernel attributed to a specific qubit, $i$, we count how many times a given feature, $j$, was re-uploaded in the past light-cone of the measured qubit, $w_i(x_j)$ (see the caption of Fig. \ref{fig:circ}).
Now, the importance score of the feature $j$ can be calculated as \cite{suzuki2024light}
\begin{equation}\label{eq:importance_score}
    P_j = \frac{1}{\mathcal{N}} \sum_{i=1}^n w_i(x_j) \lambda_i,
\end{equation}
where $\mathcal{N}$ is the normalization constant. To conclude, the features with high values of $P$ are the ones that were re-uploaded many times in the most influential local kernels.

\subsection{Methane detection using machine learning}
\label{sec:methane_detection}

We used a support vector machine classifier as a tool for methane detection and localization. We considered and compared three different kernels exploited in this classifier:
\begin{itemize}
    \item SVM$_{\rm L}$---the linear kernel is the simplest type of kernel, which assumes a linear separation of data. In our case, the linear kernel allowed for fast and efficient data processing, but its effectiveness is commonly limited in the case of more complex patterns.
    \item SVM$_{\rm RBF}$---the radial basis function kernel is more flexible than the linear kernel because it can model nonlinear relationships in the data. The use of the RBF kernel often significantly improves the generalization capabilities of the classifier, hence this kernel is commonly considered to be the ''first choice'' in the field~\cite{Nalepa2019}.
    \item SVM$_{\rm Q}$---the quantum kernel, is a simple local quantum kernel (Eq. \ref{eq:kernel}) obtained from one-qubit density matrices, $\kappa_i(x,y) = T\left[ \rho^i_x \rho^i_y \right]$, where $\rho^i_x$ is the one-qubit density matrix of the quantum embedded datapoint $x$ with the embedding circuit presented in Fig. \ref{fig:circ}. Blocks $V(x_i)$, $U(\Theta_{k,l})$ are defined in the same way as in \cite{suzuki2024light}. The $\Theta$ parameters can be optimized by maximizing centered alignment, however, such procedure is known to suffer from many problems \cite{miroszewski2023optimizing}.
\end{itemize}

\section{Results}

The validation of our methods was conducted using the unseen test set $\TestSet'$ and standard metrics such as accuracy (\ACC), sensitivity (\SEN), specificity (\SPE), F-score (\Fscore), and the Matthews correlation coefficient (\MCC). The values for \ACC, \SEN, \SPE, and \Fscore\,range from 0 to 1, with 1 indicating a perfect match with the ground truth (\GT) labels. The values of \MCC\ range from $-1$ (strong negative relationship between GT labels and predictions) to 1 (strong positive relationship). These metrics provide a comprehensive evaluation of the model's performance, considering both accuracy and the balance between different types of errors. The results were statistically analyzed.
Each model was compared with \GT\,using the Chi-square test for categorical data, and the effect size was measured using the phi coefficient ($\Phi$), which indicates the strength of the association between two categorical variables.


The experimental results obtained for all investigated kernel methods and feature subsets are gathered in Table~\ref{tab:results}. In Fig.~\ref{fig:combined_results_original}, we render example visual results of the investigated models, together with the corresponding ground-truth methane maps and elaborated superpixels which undergo classification.

In Table \ref{tab:FS}, we present the analysis of the importance of the features for the SVM$_{\rm Q}$ classifier. The two average measures of importance were used, the averaging for a specific feature was performed over all trials in which this feature was used. The importance measures are: the importance score defined in Eq. \ref{eq:importance_score} which is obtained during the training process and the sum of the \SEN, \SPE, \ACC, \Fscore, \MCC\, metrics which are obtained over the unseen test set $\TestSet'$.

\newcommand{\mywidth}{0.694}

\begin{figure*}[ht!]
\centering
\scalebox{\mywidth}{
\scriptsize
\centering
    {
    \renewcommand{\tabcolsep}{0.25mm}
    \renewcommand{\arraystretch}{0.25}
    \begin{tabular}{cccccccccc}
    \Xhline{2\arrayrulewidth}
    & (a) RGB image & & (b) Superpixel image & & (c) SVM$_{\rm L}$ & & (d) SVM$_{\rm RBF}$ & & (e) SVM$_{\rm Q}$  \\ \cline{2-4} \cline{6-6} \cline{8-8} \cline{10-10}\\
    

    \rotatebox{90}{Methane pixels: 17069 (6.51\%)} 
    &\includegraphics[width=0.25\textwidth]{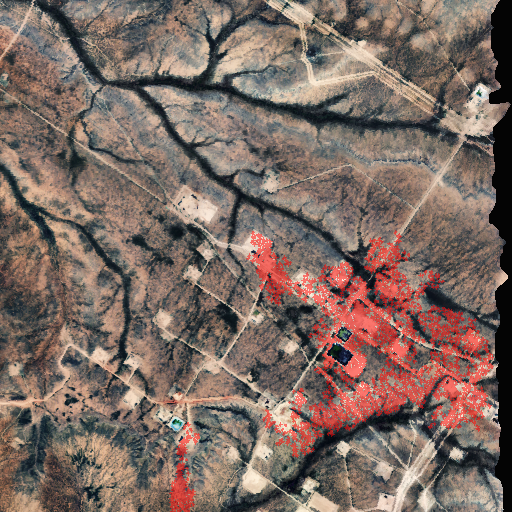} 
    &\rotatebox{90}{Superpixel count: 461} 
    &\includegraphics[width=0.25\textwidth]{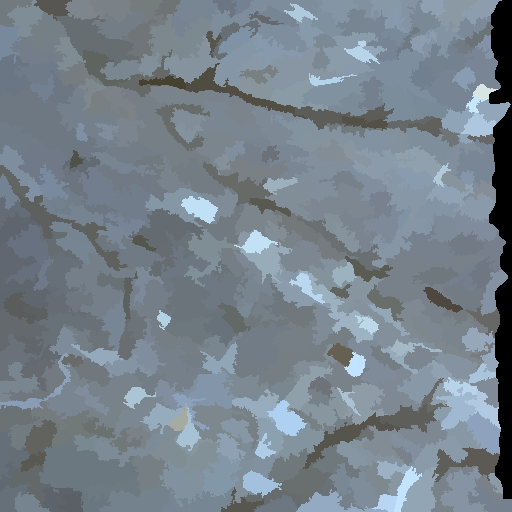} 
    &\rotatebox{90}{\TP =39, \FP =155, \FN =55, \TN =212} 
    &\includegraphics[width=0.25\textwidth]{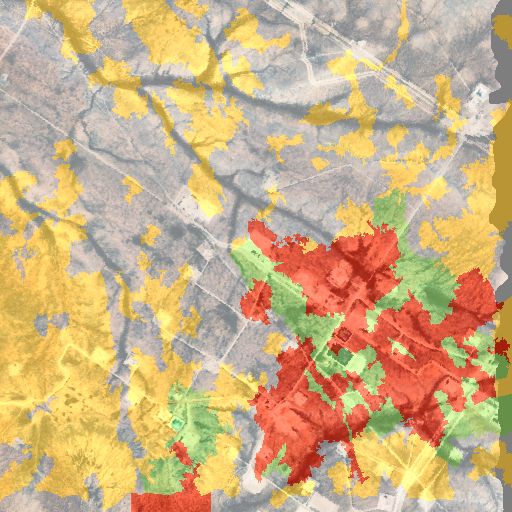} 
    &\rotatebox{90}{\TP =67, \FP =274, \FN =27, \TN =93}
    &\includegraphics[width=0.25\textwidth]{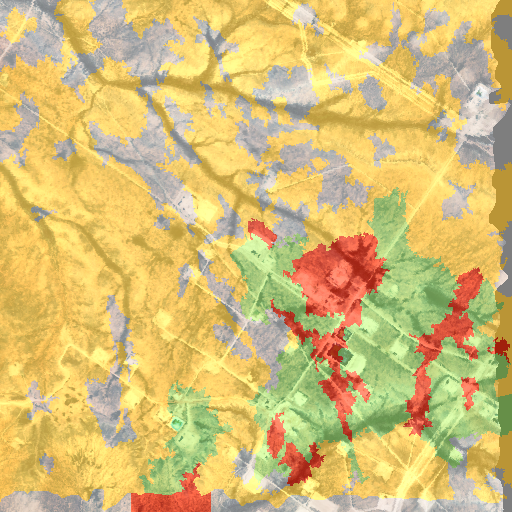} 
    &\rotatebox{90}{\TP =56, \FP =173, \FN =38, \TN =194}
    &\includegraphics[width=0.25\textwidth]{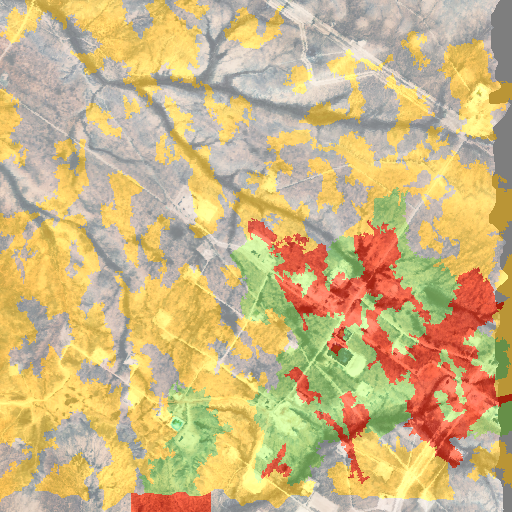} \vspace{0.25cm} \\

    \rotatebox{90}{Methane pixels: 2426 (0.93\%)} 
    &\includegraphics[width=0.25\textwidth]{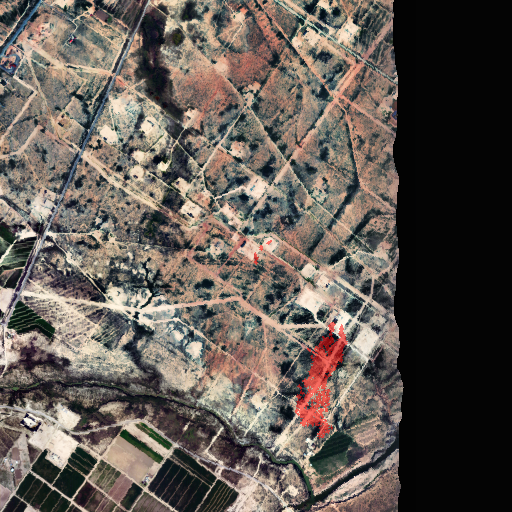} 
    &\rotatebox{90}{Superpixel count: 388} 
    &\includegraphics[width=0.25\textwidth]{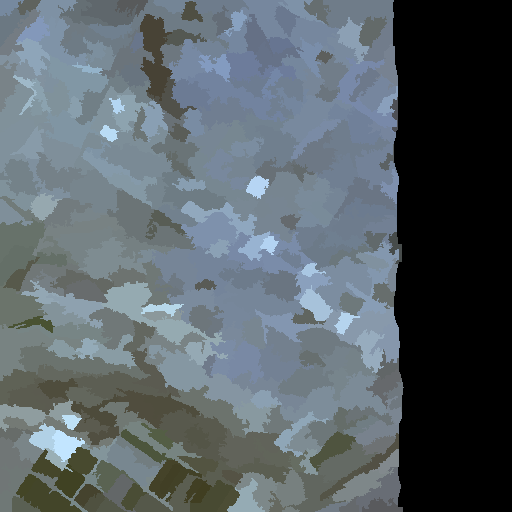} 
    &\rotatebox{90}{\TP =10, \FP =157, \FN =7, \TN =214} 
    &\includegraphics[width=0.25\textwidth]{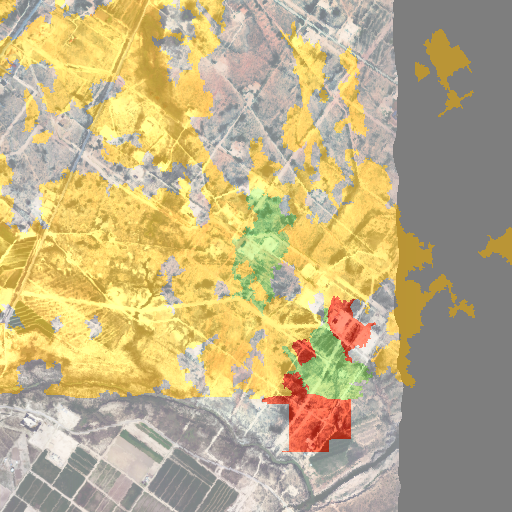} 
    &\rotatebox{90}{\TP =9, \FP =238, \FN =8, \TN =133}
    &\includegraphics[width=0.25\textwidth]{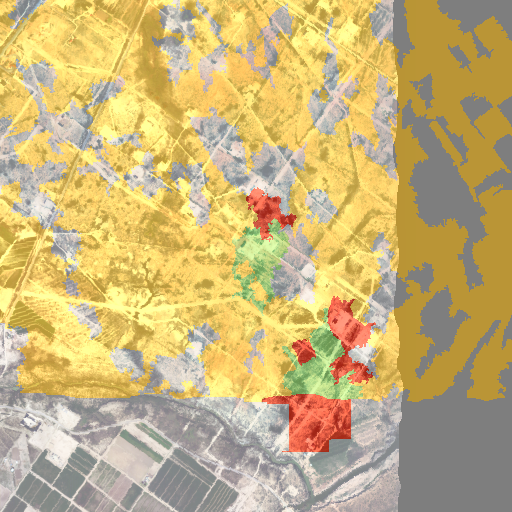} 
    &\rotatebox{90}{\TP =8, \FP =137, \FN =9, \TN =234}
    &\includegraphics[width=0.25\textwidth]{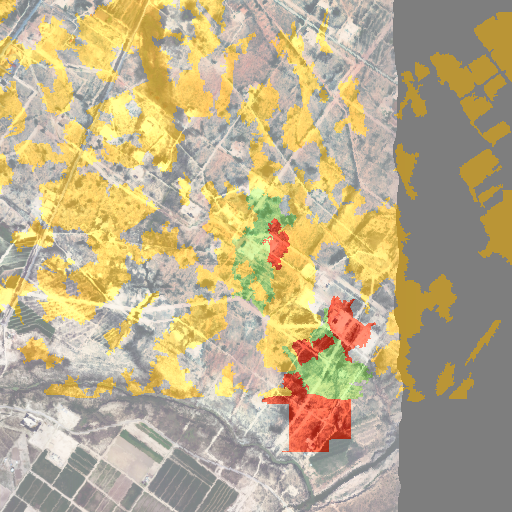} \\


    \Xhline{2\arrayrulewidth}
    \end{tabular}
    }}
\caption{Example methane detection results with the quantitative metrics obtained our approach for the original STARCOP images: (a)~RGB images with \GT\,marked in red, (b)~superpixels, (c-e)~various prediction methods (\TP---green, \FP---yellow, \FN---red).}\label{fig:combined_results_original}
\end{figure*}

\newcommand{\Feature}{\mathcal{F}}

\begin{table}[ht!]
    \centering
    \renewcommand{\tabcolsep}{2.5mm}
    \caption{The results obtained using all models with removed one feature from the feature vector ($\Feature$ denotes its identifier), without and with the mag1c enhancement map (\xmark\,and \cmark, respectively).}
    \label{tab:results}
    \scalebox{0.72}{
    \begin{tabular}{rccccccccc}
    \Xhline{2\arrayrulewidth}
    Model & mag1c & $\Feature$ & \SEN & \SPE & \ACC & \Fscore & \MCC & $p$-value & $\Phi$ 
    \\
    \hline
    
    SVM$_{\rm L}$ & \xmark & 0 & 0.617 & \cellcolor{celllightgreen}0.578 & \cellcolor{celllightgreen}0.599 & 0.622 & \cellcolor{celllightgreen}0.195 & {\color{textgreen} 0.606} & ---  \\
    SVM$_{\rm RBF}$ & \xmark & 0 & \cellcolor{celllightgreen}0.805 & 0.301 & 0.571 & \cellcolor{celllightgreen}0.668 & 0.123 & {\color{textred}0.000} & 0.05  \\
    SVM$_{\rm Q}$ & \xmark & 0 & 0.517 & 0.566 & 0.540 & 0.546 & 0.083 & {\color{textred}0.002} & 0.00\\ \hline
    
    SVM$_{\rm L}$ & \cmark & 1 & \cellcolor{celllightgreen}0.579 & 0.636 & \cellcolor{celllightgreen}0.606 & \cellcolor{celllightgreen}0.611 & \cellcolor{celllightgreen}0.215 & {\color{textred}0.002} & 0.00 \\
    SVM$_{\rm RBF}$ & \cmark & 1 & 0.373 & \cellcolor{celllightgreen}0.676 & 0.514 & 0.451 & 0.051 & {\color{textred}0.000} & 0.03  \\
    SVM$_{\rm Q}$ & \cmark & 1 & 0.517 & 0.55 & 0.533 & 0.542 & 0.067 & {\color{textred}0.006} & 0.00 \\
    \hline

    SVM$_{\rm L}$ & \cmark & 2 & 0.536 & \cellcolor{celllightgreen}0.677 & \cellcolor{celllightgreen}0.602 & 0.591 & \cellcolor{celllightgreen}0.215 & {\color{textred}0.000} & 0.01 \\
    SVM$_{\rm RBF}$ & \cmark & 2 & \cellcolor{celllightgreen}\textbf{0.813} & 0.288 & 0.569 & \cellcolor{celllightgreen}\textbf{0.669} & 0.118 & {\color{textred}0.000} & 0.06 \\
    SVM$_{\rm Q}$ & \cmark & 2 & 0.666 & 0.451 & 0.566 & 0.622 & 0.12 & {\color{textred}0.000} & 0.01  \\ \hline
    
    SVM$_{\rm L}$ & \cmark & 3 & 0.512 & \cellcolor{celllightgreen}0.676 & 0.588 & 0.571 & \cellcolor{celllightgreen}0.190 & {\color{textred}0.000} & 0.01  \\
    SVM$_{\rm RBF}$ & \cmark & 3 & 0.551 & 0.564 & 0.557 & 0.571 & 0.114 & {\color{textgreen} 0.039} & 0.00 \\
    SVM$_{\rm Q}$ & \cmark & 3 & \cellcolor{celllightgreen}0.692 & 0.486 & \cellcolor{celllightgreen}0.596 & \cellcolor{celllightgreen}0.647 & 0.182 & {\color{textred}0.000} & 0.01  \\ \hline

    SVM$_{\rm L}$ & \cmark & 4 & 0.621 & \cellcolor{celllightgreen}0.580 & \cellcolor{celllightgreen}0.602 & 0.625 & \cellcolor{celllightgreen}0.200 & {\color{textgreen} 0.658} & --- \\
    SVM$_{\rm RBF}$ & \cmark & 4 & \cellcolor{celllightgreen}0.805 & 0.299 & 0.570 & \cellcolor{celllightgreen}0.668 & 0.122 & {\color{textred}0.000} & 0.05 \\
    SVM$_{\rm Q}$ & \cmark & 4 & 0.427 & 0.565 & 0.491 & 0.474 & -0.008 & {\color{textred}0.000} & 0.01 \\ \hline
    
    SVM$_{\rm L}$ & \cmark & 5 & 0.637 & 0.562 & \cellcolor{celllightgreen}0.602 & 0.632 & \cellcolor{celllightgreen}0.199 & {\color{textgreen} 0.631} & ---  \\
    SVM$_{\rm RBF}$ & \cmark & 5 & \cellcolor{celllightgreen}0.808 & 0.285 & 0.565 & \cellcolor{celllightgreen}0.666 & 0.109 & {\color{textred}0.000} & 0.06  \\
    SVM$_{\rm Q}$ & \cmark & 5 & 0.264 & \cellcolor{celllightgreen}\textbf{0.774} & 0.501 & 0.362 & 0.044 & {\color{textred}0.000} & 0.09 \\ \hline

    SVM$_{\rm L}$ & \cmark & 6 & 0.631 & 0.571 & 0.603 & 0.630 & 0.202 & {\color{textgreen} 0.941} & ---  \\
    SVM$_{\rm RBF}$ & \cmark & 6 & \cellcolor{celllightgreen}0.808 & 0.295 & 0.570 & \cellcolor{celllightgreen}0.668 & 0.120 & {\color{textred}0.000} & 0.06  \\
    SVM$_{\rm Q}$ & \cmark & 6 & 0.623 & \cellcolor{celllightgreen}0.615 & \cellcolor{celllightgreen}\textbf{0.619} & 0.637 & \cellcolor{celllightgreen}\textbf{0.237} & {\color{textgreen} 0.210} & ---  \\ \hline
    
    SVM$_{\rm L}$ & \cmark & 7 & 0.630 & 0.572 & 0.603 & 0.630 & 0.202 & {\color{textgreen} 1.000} & ---\\
    SVM$_{\rm RBF}$ & \cmark & 7 & \cellcolor{celllightgreen}0.808 & 0.298 & 0.571 & \cellcolor{celllightgreen}0.669 & 0.123 & {\color{textred}0.000} & 0.05  \\
    SVM$_{\rm Q}$ & \cmark & 7 & 0.592 & \cellcolor{celllightgreen}0.644 & \cellcolor{celllightgreen}0.616 & 0.623 & \cellcolor{celllightgreen}0.235 & {\color{textred}0.004} & 0.00 \\ \hline
    
    SVM$_{\rm L}$ & \cmark & 8 & 0.631 & \cellcolor{celllightgreen}0.572 & \cellcolor{celllightgreen}0.604 & 0.630 & \cellcolor{celllightgreen}0.203 & {\color{textgreen} 0.971} & --- \\
    SVM$_{\rm RBF}$ & \cmark & 8 & \cellcolor{celllightgreen}0.808 & 0.296 & 0.570 & \cellcolor{celllightgreen}0.668 & 0.122 & {\color{textred}0.000} & 0.05  \\
    SVM$_{\rm Q}$ & \cmark & 8 & 0.692 & 0.486 & 0.596 & 0.647 & 0.182 & {\color{textgreen} 0.112} & --- \\ \hline

    \Xhline{2\arrayrulewidth}
    \end{tabular}}
\end{table}

\begin{table*}[ht!]
    \centering
    \renewcommand{\tabcolsep}{2.5mm}
    \caption{Mean and the standard deviation of the importance score $\langle P_j \rangle$, and a mean sum of the metrics $\langle S_j \rangle = \langle$\SEN + \SPE + \ACC + \Fscore+ \MCC$\rangle$\ obtained on the test set for the light-cone circuit for eight models trained including each feature. The three highest values of $\langle P_j \rangle$ and $\langle S_j \rangle$ are in bold.}
    \label{tab:FS}
    \scalebox{0.85}{
    \begin{tabular}{rccccccccc}
    \Xhline{2\arrayrulewidth}
    $j$ & \multicolumn{1}{c}{$6$} & \multicolumn{1}{c}{$1$} & \multicolumn{1}{c}{$3$} & \multicolumn{1}{c}{$7$} & \multicolumn{1}{c}{$2$} & \multicolumn{1}{c}{$5$} & \multicolumn{1}{c}{$4$} & \multicolumn{1}{c}{$8$} & \multicolumn{1}{c}{$0$} \\ \hline
$\langle P_j \rangle$ & $\mathbf{0.130 (31)}$ & $\mathbf{0.129 (30)}$ & $\mathbf{0.128 (10)}$ & $0.126 (23)$ & $0.126 (30)$ & $0.126 (35)$ & $0.125 (27)$ & $0.118 (21)$ & $0.118 (30)$  \\
$\langle S_j \rangle$ & $2.34 (30)$ & $\mathbf{2.40 (32)}$ & $2.35 (32)$ & $2.34 (30)$ & $2.38 (33)$ & $\mathbf{2.44 (28)}$ & $\mathbf{2.43 (28)}$  & $2.35 (32)$  & $2.39 (32)$        \\ \hline
    
    \Xhline{2\arrayrulewidth}
    \end{tabular}}
\end{table*}

\section{Discussion}

There are several insights that may be learned from the experimental results of our study. Without the mag1c enhancement map, the SVM$_{\rm L}$ model achieved an \ACC\,of 0.599, \SEN\,of 0.617, and \SPE\,of 0.578. The \Fscore\,score was 0.622, and the MCC was 0.195. The SVM$_{\rm RBF}$ model showed a higher \SEN\,of 0.805 but lower \SPE\,of 0.301, resulting in an a \ACC\,of 0.571 and an \Fscore\,of 0.668. The \MCC\,was 0.123. The SVM$_{\rm Q}$ model had a balanced performance with \SEN\,of 0.517, \SPE\,of 0.566, \ACC\,of 0.540, \Fscore\,of 0.546, and \MCC of 0.083. In this case, no significant statistical differences between the prediction and \GT\,were observed only in the case of SVM$_{\rm L}$. This means that for the remaining models it is necessary to enhance the information using, for example, the methane enhancement maps obtained using the mag1c algorithm~\cite{2020_Foote}.

With the mag1c enhancement map, the performance of the models increased. In the case of quantum methods, we had to restrict the number of features to 8 features (being one of the limitations of the technique), so we conducted experiments aimed at disabling successive bands to ``make room'' for the mag1c enhancement map. For the SVM$_{\rm L}$ model, the greatest increase in metrics was observed when replacing band 1. \ACC\,increased to 0.606, \SEN\,to 0.579, and \SPE\,to 0.636. The \Fscore\,increased to 0.611, and the \MCC\,to 0.215. However, no statistically significant differences for this model were noted when disabling higher bands (from 4 to 8). The \MCC\,value for these cases was $\geq$ 0.200. This indicates that the model's performance remains relatively stable and robust even when higher bands are disabled, suggesting that the enhancement map compensates the loss of these features. The linearity of the kernel may contribute to this stability, as linear kernels are well-suited for data that are linearly separable, making the model less sensitive to changes in the input features.
In the case of SVM$_{\rm RBF}$, replacing one of the bands with the enhancement map did not significantly affect the results compared to \GT. Of note, the performance of RBF-powered SVMs is strongly affected by the values of $\gamma$ and $C$---grid-searching these parameters might help further improve the capabilities of these models. 
In the case of SVM$_{\rm Q}$, a statistically significant impact was observed when replacing bands 6 and 8 with mag1c. The most favorable metric values for this model were obtained when replacing band 6, with \ACC = 0.603, \SEN = 0.631, and \SPE = 0.571. \MCC\,increased from 0.083 to 0.237, which is also the highest result obtained for all tested models. The advantage of the quantum solution in this case may be due to better model fitting to patterns in the data, allowing for more effective use of the information contained in the enhancement map. Finally, the visualizations rendered in Fig.~\ref{fig:combined_results_original} indicate that the models offer high-quality classification thus localization of methane areas.

For each of the models, a significant limitation is the number of observations used in training, which is due to the limited computational capabilities of quantum calculations. Nevertheless, it is important to highlight that despite using ''only'' 30 observations, SVM$_{\rm L}$ and SVM$_{\rm Q}$ showed no statistically significant differences, as confirmed by the test. Of note, the highest \MCC\,was obtained for the quantum model.

For feature selection, both measures $\langle P_j \rangle$ and $\langle S_j \rangle$ did not result in stable values, as their standard deviation is higher than the differences of those measures for different features. When considering the three most important features according to the two methods, they agree only on the importance of the feature $1$. Interestingly, none of the measures places mag1c in the most important features.

\section{Conclusions}

In this study, we presented the application of light-cone feature selection in quantum machine learning for methane detection and localization using HSIs. Experimental results showed that the quantum kernel classifier leads to the comparable classification performance as the traditional methods such as SVM with linear and RBF kernels. The light-cone feature selection, although being an intriguing theoretical construction, in this case did not produce stable and reliable results in choosing the most important features for classification. Future research can focus on the centered alignment optimization procedure, modifying the importance score and extending this technique to other areas of detection and classification.

\small
\bibliographystyle{IEEEtranN}
\bibliography{references}

\end{document}